\documentclass[amsmath,notitlepage,amssymb,aps,showkeys,floatfix,prd,a4paper,
  twocolumn,nofootinbib]{revtex4-1}

\usepackage{graphicx}
\usepackage{amsmath}
\usepackage{epstopdf}
\usepackage{amsfonts,amssymb}
\usepackage[utf8]{inputenc}
\usepackage{pstricks}
\usepackage{color}
\usepackage{hyperref}
\usepackage[toc]{appendix}

\usepackage[compat=1.0.0]{tikz-feynman}
\usepackage{color, colortbl}
\definecolor{Gray}{gray}{0.9}
\usepackage{float}
\usepackage{tikz-feynman}
\usepackage{forest}

\usepackage{verbatim}    
\usepackage{dcolumn}
\usepackage{bm}
\usepackage{epsfig}
\usepackage{braket}
\usepackage[normalem]{ulem} 
\usepackage{sidecap}

\newcommand{\be}{\begin{equation}}
\newcommand{\ee}{\end{equation}}
\newcommand{\ben}{\begin{eqnarray}}
\newcommand{\een}{\end{eqnarray}}

\newcommand{\e}{\text{e}}
\def\MeV{\mbox{ MeV}}

\def\MeV{\mbox{ MeV}} 
 
\def\TeV{\mbox{ TeV}} 
 
\newcommand{\pslash}{\not{\hbox{\kern-2.3pt $p$}}}
\newcommand{\pdslash}{\not{\hbox{\kern-2pt $\partial$}}}

\begin{document}
\title{The $\chi_{c1}(4274)$ multiplicity in heavy-ion collisions}

\author{L.M. Abreu}
\email{luciano.abreu@ufba.br}
\affiliation{ Instituto de F\'isica, Universidade Federal da Bahia,
Campus Universit\'ario de Ondina, 40170-115, Bahia, Brazil}
\affiliation{Instituto de F\'{\i}sica, Universidade de S\~{a}o Paulo, 
Rua do Mat\~ao, 1371, CEP 05508-090,  S\~{a}o Paulo, SP, Brazil}

\author{A.L.M. Britto}
\email{andrebritto@ufrb.edu.br}
\affiliation{ Centro de Ciências Exatas e Tecnológicas, Universidade Federal do Recôncavo da Bahia, R. Rui Barbosa, Cruz das Almas, 44380-000, Bahia, Brazil}\affiliation{ Instituto de F\'isica, Universidade Federal da Bahia,
Campus Universit\'ario de Ondina, 40170-115, Bahia, Brazil}

\author{F. S. Navarra}
\email{navarra@if.usp.br}
\affiliation{Instituto de F\'{\i}sica, Universidade de S\~{a}o Paulo, 
Rua do Mat\~ao, 1371, CEP 05508-090,  S\~{a}o Paulo, SP, Brazil}
\author{H. P. L. Vieira}
\email{luciano.abreu@ufba.br}
\affiliation{ Instituto de F\'isica, Universidade Federal da Bahia,
Campus Universit\'ario de Ondina, 40170-115, Bahia, Brazil}

\begin{abstract}
   
In a previous work we computed the thermally-averaged cross sections for the production and suppression reactions of the $\chi_{c1}(4274)$ state the hot hadron gas formed in heavy ion collisions.  In the present work we estimate the final yield of the $\chi_{c1}(4274)$ state in these collisions. We use the coalescence model to fix the initial multiplicity of the $\chi_{c1}(4274)$, which is treated as a $P-$wave bound state of $D_s\bar D_{s0}$ and also as a compact tetraquark.  The Bjorken picture is used to model the hydrodynamic expansion and cooling. Then, the kinetic equation is solved to evaluate the time evolution of the $\chi_{c1}(4274)$ yield during the hot hadron gas phase. Since the $\chi_{c1}(4274)$ decay width is large it might decay inside the hadron gas. Therefore we also include the $\chi_{c1}(4274)$ decay and regeneration terms by means of an effective coupling, estimated from the available data. The combined effects of hadronic interactions and the $\chi_{c1}(4274)$ decay have a strong impact on the final yield. Also, predictions of the $\chi_{c1}(4274)$ multiplicity as a function of centrality and of the charged hadron multiplicity measured at midrapidity $[dN_{ch}/d\eta \,(\eta<0.5)]$ are presented. 
Finally, we calculate the yield of a proposed $P-$wave molecular state of $D_s \bar{D_{s0}}$, $Y^{\prime}(4274)$, characterized by a smaller width and smaller coupling constant obtained from the Weinberg compositeness condition.

\end{abstract}
\maketitle

\section{INTRODUCTION}



During the last two decades, the number of new hadrons has increased substantially \cite{qwg}, with the observation of new  states whose properties are incompatible with the quark model predictions. One of these unconventional structures with very intriguing properties is the charmoniumlike $\chi_{c1}(4274)$ state.
It has been observed by the LHCb Collaboration in the amplitude analysis of the decay $B^+\rightarrow J/ \psi \phi K^+$. Its quantum numbers have been established to be $I^G(J^{PC})=0^+(1^{++})$ with statistical significance of $6.0 \, \sigma$, and its corresponding measured mass and width are~\cite{LHCb:2016axx,LHCb:2016nsl} 
\begin{align}
    M_\chi =4273.3\pm 8.3^{+17.2}_{-3.6}~\MeV,
    \Gamma_\chi =56\pm 11^{+8.0}_{-11}~\MeV,
\label{masswidthChi1}
\end{align}
at $5.8 \, \sigma$ of significance. These values of mass and width are consistent with a previous measurement claimed by the CDF collaboration~\cite{CDF:2011pep}:
\begin{align}
M_\chi=4274.4^{+8.4}_{-6.7}\MeV,~\Gamma_\chi=32.3^{+21.9}_{-15.3}\MeV.
\label{masswidthChi2}
\end{align}


Several works were dedicated to understand the properties and intrinsic nature of the $\chi_{c1}(4274)$  ~\cite{Chen:2016oma,Wang:2016dcb,Stancu:2009ka,Zhu:2016arf,Lu:2016cwr,Gui:2018rvv,Agaev:2017foq,Maiani:2016wlq,He,zhu2022possible}. There were 
attempts to interpret $\chi_{c1}(4274)$ as a $S-$wave $cs\bar{s}\bar{c}$ tetraquark state, as a conventional $\chi_{c1}(3^3 P_1)$ structure, 
as a color triplet and sextet diquark-antidiquark configuration, as a molecular state of mesons, and others (we refer the reader to Ref.~\cite{Britto:2023iso} for a more detailed discussion).

Motivated by the fact that the $\chi_{c1}(4274)$ mass is just 12 MeV below the $D_s \bar{D}_{s0}(2317)$ threshold, the author of Ref.~\cite{He} has performed an analysis of the $\chi_{c1}(4274)$ as a $P-$wave bound state of $D_s D_{s0}(2317)$ in a quasipotential Bethe-Salpeter equation approach, with a partial wave decomposition on spin parity.
The quantum numbers of $D_s $ and $\bar{D}_{s0}(2317)$ (denoted from now on by $\bar{D}_{s0}$ ) are $0^{-} $ and $ 0^{+}$ and hence the  binding mechanism between these two mesons must be in $P-$wave and strong enough to generate a bound state.  However, this idea was not supported by the results found in Ref.~\cite{zhu2022possible}, where calculations based on an effective approach with the Weinberg compositeness condition gave a much smaller partial decay width for the $\chi_{c1}(4274)$ than the one experimentally  observed. In view of these results the authors of
Ref.~\cite{zhu2022possible}  proposed the existence of a new $D_s D_{s0}$ $P-$wave bound state called $Y'(4274)$. 

The debate on the properties of the $\chi_{c1}(4274)$ will continue and certainly  there will be more contributions on this issue. From the experimental side, heavy-ion collisions (HICs) appear as a promising scenario to investigate the properties of exotic states, as was pointed out in previous works~\cite{Abreu1,Abreu2,XProd1,XProd2, Abreu:2017cof,Abreu:2018mnc}. The final  $\chi_{c1}(4274)$  multiplicity will depend on the interaction cross sections, which, in turn, depend on the spatial configuration of the quarks. Meson molecules are larger, and therefore have greater cross sections, which means that they will have a stronger interaction with the hadronic medium than compact tetraquarks. 
In Ref.~\cite{Britto:2023iso} we have studied the interactions of the  $\chi_{c1}(4274)$ with a hadronic medium via an effective approach, and we  have found significant differences between the thermally-averaged cross sections for the $\chi_{c1}(4274)$ production and  suppression reactions. These findings motivate us to pursue this investigation and use the obtained thermal cross sections to estimate the final yield of the $\chi_{c1}(4274)$ in a heavy-ion collision environment. We use the coalescence model to fix the initial multiplicity of the $\chi_{c1}(4274)$, which is treated as a $P-$wave bound state of $D_s\bar D_{s0}$ and also as a compact tetraquark. The hydrodynamic expansion is described by the Bjorken  model. With these inputs we solve the kinetic equation to determine the time evolution of the $\chi_{c1}(4274)$ yield during the hot hadron gas phase. Finally we present predictions for the $\chi_{c1}(4274)$ multiplicity as a function of centrality and charged hadron multiplicity measured at midrapidity ($[dN_{ch}/d\eta \,(\eta<0.5)]$).  Additionally,  we compute the yield of the $P-$wave molecular state of $D_s \bar{D_{s0}}$, $Y^{\prime}(4274)$ proposed in Ref.~\cite{zhu2022possible}.  

This work is organized as follows. In the next section, we present the formalism: the rate equation that drives  the time evolution of the $\chi_{c1}$ abundance, the coalescence model and how the dependence with centrality and charged hadron multiplicity measured at midrapidity is implemented. Section III is devoted to discuss the results  and in the last section we present the concluding remarks.

\section{Formalism}
\label{Formalism}


\subsection{Kinetic equation}
\label{kineticequation}

In order to estimate the time evolution of the $\chi_{c1}$ yield  we employ the integro-differential equation given by \cite{Abreu1,Abreu2,XProd1,XProd2, Abreu:2017cof,Abreu:2018mnc}:
\begin{align}
    \frac{dN_\chi(\tau)}{d\tau}=\sum_{\substack{c,c'=D_{s}D_{s0}
    \\ a=\pi,K,\eta,\\ \rho,K^*,\omega}}
    &\left[\langle \sigma_{cc'\rightarrow \chi a} v_{cc'} \rangle
    n_c(\tau)N_{c'}(\tau)\right.\notag\\-&
    \left.
    \langle \sigma_{\chi a\rightarrow cc' } v_{\chi a}\rangle
     n_\phi(\tau)N_{\chi}(\tau)
    \right]\notag\\
&
+\langle\sigma_{J/\psi \phi\rightarrow Y} v_{J/\psi \phi}\rangle n_\phi(\tau) N_{J/\psi}(\tau) \notag\\-
&\Gamma_{\chi_{c1}} N_\chi(\tau),\label{rateequation}
\end{align}
where $N_{\chi}(\tau)$ and $N_{c'}(\tau)$ represent the abundance of $\chi_{c1}$ and the charmed (strange) mesons at proper time $\tau$; $n_c(\tau)$, and $n_\phi(\tau)$ are the number densities of charmed strange mesons and light mesons, respectively. We assume that the hadron gas is in thermal equilibrium, and its constituents have their respective densities following the Boltzmann distribution, i.e.
\begin{align}
    n_i=\frac{1}{2\pi^2}\gamma_i g_im_i^2 T(\tau) K_2\left( \frac{m_i}{T(\tau)}\right),\label{statistical}
\end{align}
with $\gamma_i$, $g_i$, and $m_i$ being the fugacity, degeneracy and mass of the particle $i$, respectively; $T(\tau)$ is the time-dependent temperature. The multiplicity, $N_i(\tau)$, is obtained by multiplying the abundance, $n_i(\tau)$,  by the volume $V(\tau)$.

In Eq.~\ref{rateequation}, $\langle \sigma_{i \rightarrow \chi f }v_{i} \rangle$ are the thermally averaged cross section calculated and discussed in Ref.~\cite{Britto:2023iso}.

The decay width of $\chi_{c1}$ ($\Gamma_{\chi_{c1}}$) is relatively large and its lifetime may be shorter than the lifetime of the hadron gas phase assumed in this work (of the order of 10 fm/c). Therefore the decay of $\chi_{c1}$ as well as its regeneration from the daughter particles of the possible processes are included in the last two lines of Eq.~\ref{rateequation}. According to the Review of Particle Physics (RPP) 2022~\cite{Workman:2022ynf}, only the reaction $ \chi_{c1} \rightarrow J/\psi \phi $ has been seen, and hence only this one is considered here to estimate the mentioned decay and regeneration terms. We will proceed as in  Refs.~\cite{Cho:2015qca,Abreu:2020ony}. The  decay width   $\Gamma_{\chi_{c1}}$ is determined with the help of the effective Lagrangian
\begin{align}
\mathcal{L}=g_{\chi_{c1} J/\psi \phi} \, \epsilon_{\mu\nu\rho\sigma} \, \phi^\mu \, \psi^\nu \, (\partial^\rho \chi^\sigma - \partial^\sigma \chi^\rho ),
\label{vertex}
\end{align}
with the coupling constant $g_{\chi_{c1} J/\psi \phi}$ being fixed by fitting the experimental data from \cite{LHCb:2016axx,LHCb:2016nsl}. The  decay rate can be written as
\begin{align}
\Gamma_{\chi_{c1}}(\sqrt{s})=	\frac{p_{cm}(\sqrt{s})}{8\pi s g_\chi}|\mathcal{M}|^2,
\label{decayrate1}
\end{align}
where $p_{cm}$ is the three-momentum in the center-of-mass frame; $\mathcal{M}$ is the tree-level amplitude of the decay rate expressed by:
\begin{align}
 M= g_{\chi_{c1} J/\psi \phi} \epsilon^{\mu\nu\rho\sigma}\varepsilon_\mu^*(p_2)\varepsilon_\nu^*(p_3)
 \left[
 p_{1_\rho}\varepsilon_\sigma(p_1)-
 p_{1_\sigma}\varepsilon_\rho(p_1)
 \right],
\label{ampl}
\end{align}
with the momenta of the states $\chi_{c1}$, $J/\psi$ and $\phi$ being given by $p_1$, $p_2$ and $p_3$, respectively. The polarization vector is given by $\varepsilon(p)$. Thus, using the experimental value of the $\chi_{c1}-$decay width in Eq.~(\ref{masswidthChi1}), we fix the value of the coupling constant as $g_{\chi_{c1} J/\psi \phi}=0.56\pm 0.07$.

Also, the cross section for the regeneration process is assumed to be given by the spin-averaged relativistic Breit-Wigner approximation:
\begin{align}
    \sigma(\sqrt{s})=\frac{g_\chi}{ g_\psi g_\phi }\frac{4\pi}{p_{cm}^2}\frac{s\Gamma_{\chi_{c1}}^2(\sqrt{s})}{(s-M_\chi^2)+s\Gamma_{\chi_{c1}}^2(\sqrt{s})},
\end{align}
where $g_\chi$, $g_\psi$ and $g_\phi$ are the degeneracy of the $\chi_{c1}$, $J/\psi$ and $\phi$ respectively. The thermally-averaged cross section, 
$\langle\sigma_{J/\psi \phi\rightarrow \chi} v_{J/\psi}\rangle$, is therefore calculated in the same way as done in~\cite{Britto:2023iso}. Additionally, since the decay width $\Gamma$ averaged over the thermal distribution does not present a strong dependence on the temperature, 
we use its value in the vacuum.

Finally, it is worth noticing that the coupling constant $g_{\chi D_s D_{s0}}$ has been determined theoretically in Ref.~\cite{zhu2022possible} by means of the Weinberg compositeness condition~\cite{weinberg,salam}, assuming $\chi_{c1}$ as a $P-$wave molecular state of $D_s \bar{D}_{s0}$. With this formalism the predicted decay width is $\Gamma_{\chi_{c1}}^{(Theo)} \sim 1.46\MeV$, much smaller than the experimental one ($\Gamma_{\chi_{c1}}^{(Exp)}$) given in Eq.~(\ref{masswidthChi1}). This point has been used by the authors of~\cite{zhu2022possible} to argue that the molecular interpretation of $\chi_{c1}$ is disfavored, as well as to suggest that it would be possible to observe a $P-$wave molecular state of $D_s \bar{D}_{s0}$ (so-called $Y^{\prime}(4274)$) at Belle or Belle II experiments. We employ here a different strategy: in order to obtain predictions for observables of our interest based on existing experimental data, we assume that relevant contributions for the observed $\Gamma_{\chi_{c1}}$ are effectively encoded in the coupling $g_{\chi_{c1} J/\psi \phi}$, discussed above from Eq.~(\ref{vertex}) on, and that the mechanisms involving the coupling among $\chi_{c1}, D_s $ and $ D_{s0}$ have a relevant participation. 
In this sense, we remark that the reaction $\chi_{c1} \rightarrow   J/ \psi \phi $ has been evaluated in Ref.~\cite{zhu2022possible} from the triangular diagrams $\chi_{c1} \rightarrow D_s D_{s0}(\rightarrow J/ \psi D_s^{\star}) \rightarrow  J/ \psi \phi $ and $\chi_{c1} \rightarrow D_s D_{s0}(\rightarrow \phi D_s^{\star}) \rightarrow  \phi J/ \psi $. Then, the decay rate coming from these diagrams is proportional to $g_{\chi D_s D_{s0}}^2$ while the $\Gamma_{\chi_{c1}}$ calculated through Eq.~(\ref{decayrate1}) is proportional to $g_{\chi_{c1} J/\psi \phi}^2$. Therefore, to reproduce the experimental decay width we use the prescription:
\begin{align}
   g_{\chi D_s D_{s0}} \rightarrow \Tilde{g}_{\chi D_s D_{s0}} = \sqrt{A} g_{\chi D_s D_{s0}}, 
   \label{rescaling}
\end{align}
where $A = \Gamma_{\chi_{c1}}^{(Exp)} / \Gamma_{\chi_{c1}}^{(Theo)}$. This might be interpreted as follows: a factor $\sqrt{A}$ with the rate between the effective coupling estimated via Eq.~(\ref{decayrate1}) and the theoretical coupling constant $g_{\chi D_s D_{s0}}$ is introduced in $\chi ,  D_s ,  D_{s0}$ coupling to make the theoretical prediction in accordance with data.  
Then, using the values for $\Gamma_{\chi_{c1}}^{(Exp)} $, $ \Gamma_{\chi_{c1}}^{(Theo)}$ and $g_{\chi D_s D_{s0}}$ in~\cite{zhu2022possible,Britto:2023iso}, we get $\Tilde{g}_{\chi D_s D_{s0}}=81\pm 16$. 
Hence, unless otherwise stated, in the calculations we use the thermally-averaged cross sections defined in Ref.~\cite{Britto:2023iso}, but replacing $g_{\chi D_s D_{s0}} $ according to Eq.~(\ref{rescaling}). This produces an increase in the magnitude of the thermally-averaged cross sections without qualitative modifications. Another consequence, at a more profound level, is that this prescription makes the nature of the coupling used here independent of a particular interpretation, since it no longer obeys the Weinberg compositeness condition.  The intrinsic nature of the $\chi_{c1}$ state will be given by  the initial condition employed (see discussion in a subsequent subsection).



\subsection{Hydrodynamic expansion}

The hadron gas evolution is modeled by the boost invariant Bjorken picture with an accelerated transverse expansion, in which the volume and temperature as a function of the proper time $\tau$ are as follows\cite{Abreu1, Abreu2, Abreu:2017cof, Abreu:2018mnc}:
\begin{align}
    V(\tau)&=\pi\left[R_C+v_C(\tau-\tau_C)+\frac{a_C}{2}(\tau-\tau_C)^2)\right]^2 c \tau, \notag\\
    T(\tau)&=T_C-(T_H-T_F)\left(\frac{\tau-\tau_H}{\tau_F-\tau_H}\right)^{\frac{4}{5}},\label{bjorken}
\end{align}
where $R_C, \upsilon_C, a_C$ and  $T_C$ denote the transverse size, transverse velocity, transverse acceleration and temperature at the chemical freeze-out time $\tau_C$, respectively; $T_H (T_F)$ is the temperature at the end of the mixed phase (kinetic freeze-out) time $\tau_H (\tau_F)$. 
The parameters in Eq. (\ref{bjorken}) are fixed according to Ref.~\cite{EXHIC}, for a hadronic medium formed in central Pb-Pb collisions at $\sqrt{s_{NN}} = 5.02$ TeV, and the set of parameters is given in Table~\ref{tabela1}. 

    \begin{table}[h!]
       \caption{In the first three lines we list the parameters used in Eq.~(\ref{bjorken}) for the hydrodynamic expansion of the hadronic medium produced in central Pb-Pb collisions at $\sqrt{s_{NN}} = 5.02$ TeV~\cite{EXHIC}. In the next three lines we list the multiplicities of the mesons used in the model. In the last two lines we show the quark masses and multiplicities, and the frequency 
       used in the coalescence model.}
\centering \begin{tabular}{ccc}
\hline \hline $v_C(\mathrm{c})$ & $a_C\left(\mathrm{c}^2 / \mathrm{fm}\right)$ & $R_C(\mathrm{fm})$ \\
0.5 & 0.09 & 11 \\
\hline $\tau_C(\mathrm{fm} / \mathrm{c}) $&$ \tau_H(\mathrm{fm} / \mathrm{c}) $&$ \tau_F(\mathrm{fm} / \mathrm{c})$ \\
7.1 & 10.2 & 21.5 \\
\hline $T_C(\mathrm{MeV}) $&$ T_H(\mathrm{MeV}) $&$ T_F(\mathrm{MeV})$ \\
156 & 156 & 115 \\
\hline $N_\pi\left(\tau_H\right) $&$ N_K\left(\tau_H\right) $&$ N_\eta\left(\tau_H\right)$ \\
713 & 133 & 53 \\
\hline $N_\rho(\tau_H)$&$ N_\omega(\tau_H) $&$ N_{K^*}(\tau_H)$ \\
 183 & 53 & 61\\
\hline $N_{D_s}$ & $N_{D_{s0}}$ &  \\ 1.31 & 0.18 &  \\
\hline $m_c \ [\mathrm{MeV}]$ & $m_s \ [\mathrm{MeV}]$ & $m_q \ [\mathrm{MeV}]$ \\   
1500  & 500 & 350 \\  
\hline $N_c$ & $N_s$ & $\omega_c[\mathrm{MeV}]$ \\ 14 & 386 & 220 \\
\hline \hline \label{tabela1}
\end{tabular}
    \end{table}



    In addition, in Table~\ref{tabela1} the multiplicities of the light mesons and quarks in charmed mesons are also shown. In the case of light mesons, the fugacities in Eq.~(\ref{statistical}) are normalization parameters to adjust the multiplicities given in Table~\ref{tabela1}. On the other hand, since the charm quarks are produced in the early stages of the collision, the total number of charm quarks $(N_c)$ in charmed hadrons is assumed to be approximately conserved during the hydrodynamic expansion, i.e. $ N_c = n_c (\tau) \times V(\tau) = const.$ and this  implies a time-dependent charm-quark-fugacity factor $\gamma _c \equiv \gamma _c (\tau)$. 

\subsection{Initial conditions}

As the rate equation (\ref{rateequation}) describes the time evolution of the yield $\chi_{c1}(4274)$ from the end of the mixed phase on, we need to fix its initial conditions. To this end, we employ the coalescence model, which determines  the multiplicity of the hadronic state by overlapping the density matrix of its constituents with its Wigner Function.
Accordingly, it encodes information concerning the intrinsic structure of the system, such as angular momentum and the type and number of constituent quarks.
Thus, by using the definition of the coalescence model and after some manipulations, the number of $\chi_{c1}(4274)$ at time $\tau_c$ can be written as~\cite{EXHIC}:
\begin{align}
    N_{\chi_{c1}} \approx  & g_{\chi} \prod_{j=1}^{n}\frac{N}{g_i}
    \prod_{i=1}^{n-1}\frac{(4\pi\sigma_i^2)^{\frac{3}{2}}}{V(1+2\mu_iT\sigma_i^2)}\notag\\
    & \times\left[
    \frac{4\mu_i T\sigma_i^2}{3(1+2\mu_i T\sigma_i^2)}
    \right]^{l_i},
\label{coalmodel}
\end{align}
where $g_j$ and $N_j$ are the degeneracy and number of the $j$-th constituent of the $\chi_{c1}(4274)$, and $\sigma_i=(\mu_i\omega)^{-1/2}$; the parameter $\omega$ is the so-called oscillator frequency, assuming that the hadron internal structure is represented by an harmonic oscillator; 
the reduced mass $\mu$ is given by 
$$\mu_i^{-1}=m_{i+1}^{-1}+\left(\displaystyle\sum_{j=1}^i m_{j}\right)^{-1}
$$. 
The angular momentum $l_i$ takes on values of 0 and 1 for $S$-wave and $P$-wave, respectively.

In the present approach we explore two  possible configurations of  the $\chi_{c1}(4274)$:   a $P-$wave bound state of $D_s \bar{D}_{s0}$ and a compact tetraquark $c s \bar{c} \bar{s} $. In the case of the $P-$wave molecular state, $l = 1$ in Eq.~(\ref{coalmodel});  the angular frequency is $\omega=6B$, where $B=m_{D_s}+m_{D_{s0}}-m_{\chi_{c1}}$ is the binding energy; the number of charmed strange mesons is given in Table~\ref{tabela1}. For the compact tetraquark configuration, the frequency, the quark number and masses are summarized also in Table~\ref{tabela1}. Hence, putting all these ingredients in Eq.~(\ref{coalmodel}), the initial $\chi_{c1}(4274)$ multiplicity in the molecular and tetraquark configurations are
\begin{align}
    N_{\chi_{c1}}^{(Mol)}(\tau_H)=1.96\times 10^{-4}, \notag \\
    N_{\chi_{c1}}^{(4q)}(\tau_H)=8.27\times 10^{-6}
    \label{initialconditions}
\end{align}
As expected, $N^{(Mol)}(\tau_H) > N^{(4q)}(\tau_H)$ because a shallow bound state is easier to be formed than a compact tetraquark. In the particular case of $\chi_{c1}$, at the end of mixed phase the coalescence mechanism yields more molecules than tetraquarks by a factor of about 24. 


\subsection{System Size dependence}

The multiplicity $N_{\chi_{c1}}$ can also be expressed as a function of the charged-particle pseudorapidity density at mid-rapidity, $[dN_{ch}/d\eta \, (|\eta|<0.5)]$, which is a measurable quantity, and can be associated to the freeze-out temperature through the empirical formula~\cite{Abreu1,Abreu2,LeRoux:2021adw}:
\begin{align}
    T_F=T_{F0}\e^{-b\mathcal{N}} , 
\end{align}
where $T_{F0}=132.5\MeV$, $b=0.02$, and $\mathcal{N} \equiv [dN_{ch}/d\eta \,(|\eta|<0.5)]^{1/3}$. Assuming that the hadron gas undergoes a Bjorken-type cooling, then the freeze-out time $\tau_F$ and the freeze-out temperature are related through~\cite{Abreu1,Abreu2,LeRoux:2021adw}:
\begin{align}
    \tau_F=\tau_H\left( \frac{T_H}{T_{F0}}\right)^3\e^{3b\mathcal{N}}.
    \label{temperaturerelation}
\end{align}
Thus, $\mathcal{N}$ can be seen as an indirect measurement of the duration of the hadronic phase:  a larger system (with a larger mass number $A$) yields a larger charged-particle pseudorapidity density (bigger $\mathcal{N}$), which in its turn generates a longer hadron phase 
(bigger $\tau_F$).
So, the use of Eq.~(\ref{temperaturerelation}) in (\ref{rateequation}) allows us to indirectly estimate the dependence of $N_{\chi_{c1}}$ on the  system size.

 

In addition, following Ref.~\cite{Abreu1}, $\mathcal{N}$ can also be related to other observables like the center-of-mass energy $\sqrt{s}$ and the centrality of the collision. For a  Pb-Pb collision, the empirical formula connecting $\mathcal{N}$ with $\sqrt{s}$ is given by:
\begin{align}
    \frac{d{N}_{ch}}{d\eta}=    -2332.12+491.69\log(220.06+\sqrt{s}),
    \label{energy}
\end{align}
 while for $\mathcal{N}$ with the centrality (denoted as $x$, in \%) the relation is:
\begin{align}
\left.\frac{d{N}_{ch}}{d\eta}\right|_{|\eta|<0.5}&=2142.16-
85.76x+1.89x^2-0.03x^3+
\notag\\
&+3.67\times10^{-5}x^4-2.24\times 10^{-6}x^5+\notag\\
&+5.25\times10^{-9}x^6. 
\label{centrality}
\end{align}

\section{Results}


Now we present and discuss the results obtained by solving the equation (\ref{rateequation}) using the initial conditions given by Eq. (\ref{initialconditions}). In the plots shown below, the bands denote the uncertainties coming from the values of the coupling constants (see the discussion in section \ref{kineticequation}).

\begin{figure}[h!]
    \includegraphics[scale=.95]{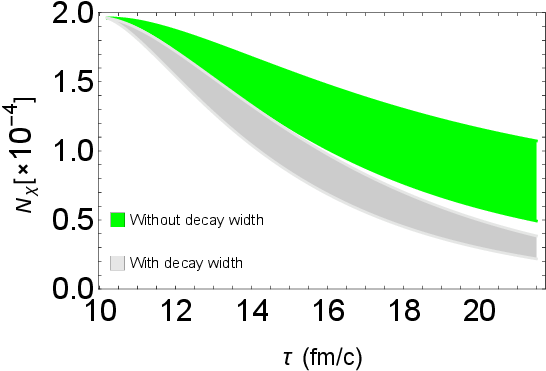}
    \caption{$\chi_{c1}$ multiplicity as a function of the proper time in central Pb-Pb collisions at $\sqrt{s_{NN}}= 5.02 \TeV$, with initial condition associated to the molecular interpretation given by Eq.~(\ref{initialconditions}). }
    \label{TimeEvNY}
\end{figure}

In Fig.~\ref{TimeEvNY} we show the time evolution of the $\chi_{c1}(4274)$ multiplicity in the case of molecular interpretation, encoded in the initial condition $N_{\chi_{c1}}^{(Mol)}(\tau_H)$ given by Eq.~(\ref{initialconditions}). For the sake of comparison, we have included the green and gray bands representing respectively the solutions $N_{\chi_{c1}}^{(Mol)} (\tau)$ of the rate equation (\eqref{rateequation}) without and with the last two lines, which are associated to the inclusion of the $\chi_{c1}$ decay rate and regeneration terms. As it can be seen from the green band, the loss terms are dominant with respect to the gain ones, and this leads to a sizable suppression of $N_{\chi_{c1}}^{(Mol)} (\tau)$, which at freeze-out time is about $40\%$ of the initial yield, taking into account the central value. When the decay rate is added, $N_{\chi_{c1}}^{(Mol)} (\tau)$ decreases faster and the  final yield is about $15\%$ of the initial one.

\begin{figure}[h!]
    \includegraphics[scale=.95]{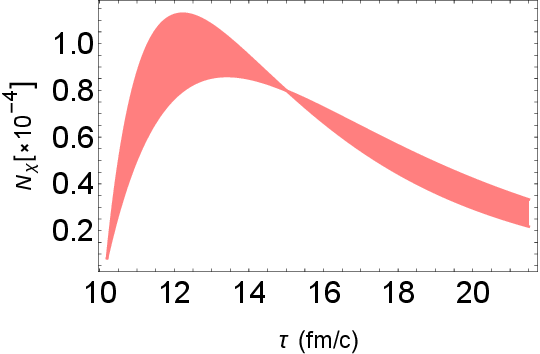}
    \caption{$\chi_{c1}$ multiplicity as a function of the proper time in central Pb-Pb collisions at $\sqrt{s_{NN}}= 5.02 \TeV$, with initial condition associated to the compact tetraquark  interpretation given by Eq.~(\ref{initialconditions}). }
    \label{TimeEvNY4q}
\end{figure}

Moving on to the compact tetraquark configuration, in Fig.~\ref{TimeEvNY4q} we present the time evolution of the $\chi_{c1}$ yield, with the initial condition  $N_{\chi_{c1}}^{(4q)}(\tau_H)$ given by Eq.~(\ref{initialconditions}); the $\chi_{c1}$ decay rate and regeneration terms have been included in the calculations.  In this case the behavior of $N_{\chi_{c1}}^{(4q)}(\tau)$ is different: at beginning of the hadron gas phase, the gain terms dominate and generate an expressive increase of the multiplicity. But as time goes on, the hadron gas expands and cools down, affecting mostly the number and densities of the conventional mesons, and the $\chi_{c1}$ production rate becomes smaller, causing a decreasing of the abundance. Notwithstanding, at the freeze-out time the $\chi_{c1}$ final yield is three times larger than the initial value. 
Accidentally,  the final $N_{\chi_{c1}}$ after the hadronic interactions is approximately of the order  $3\times 10^{-5}$ for both  molecular and compact tetraquark configurations.


\begin{figure}[h!]
    \includegraphics[scale=0.95]{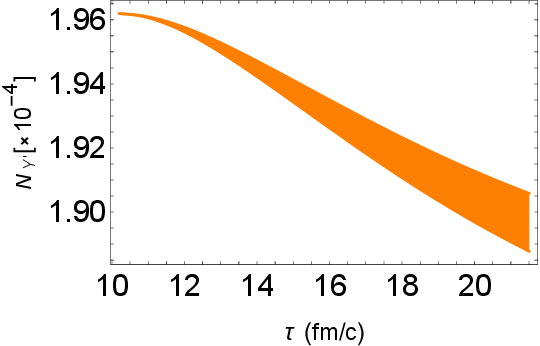}
    \caption{$Y'(4274)$ multiplicity as a function of the proper time in central Pb-Pb collisions at $\sqrt{s_{NN}}= 5.02 \TeV$, with initial condition associated to the molecular interpretation given by Eq.~(\ref{initialconditions}).}
    \label{TimeEvNYCh2}
\end{figure}

For completeness, we also consider  the state $Y'(4274)$ proposed in Ref.~\cite{zhu2022possible}, assumed there to be a $P-$wave bound state of $D_s\bar D_{s0}$, with a small width and the coupling constant obtained via Weinberg compositeness condition.   
We use the initial condition $N_{\chi_{c1}}^{(Mol)}(\tau_H)$ given by Eq.~(\ref{initialconditions}), and neglect the decay and regeneration terms since the small width implies a decay time  which is longer than the duration of the hadron gas phase. 
In Fig.~\ref{TimeEvNYCh2} we plot the $Y'(4274)$ multiplicity as a function of the proper time. The numbers in the figure suggest that the final yield undergoes a small suppression, of the order of $3\%$, which means that the gain and loss terms in Eq.~(\ref{rateequation}) contribute approximately equally, making the $Y'(4274)$ multiplicity almost constant. As we can see, there is a clear distinction between the predicted  molecular state $Y'(4274)$ (obeying the Weinberg compositeness condition) and the $\chi_{c1}$ analyzed previously.  The final yield of the 
latter is almost one order of magnitude smaller than that of the former.

\begin{figure}[h!]
\includegraphics[scale=0.95]{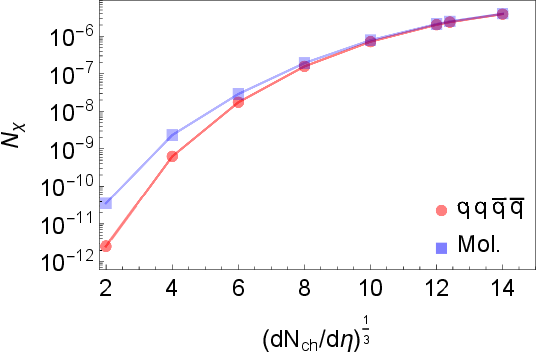}
    \caption{$\chi_{c1}$ multiplicity as a function of $\mathcal{N}$. The curves represent the results obtained with initial conditions 
    associated to the molecular and compact tetraquark configurations given by Eq.~(\ref{initialconditions}).}
    \label{NYdNdEta}
\end{figure}

In Fig.~\ref{NYdNdEta} we show the $\chi_{c1}$ multiplicity as a function of $\mathcal{N}$. As expected, it grows with the size of the system. Heavy-ion collision experiments can potentially observe a  much larger number of $\chi_{c1}$'s (larger systems like Pb-Pb systems are characterized by $\mathcal{N}\sim 10-12.5$) and may provide a very interesting environment for the study of their properties. Curiously, the curves of the molecular and compact tetraquark configurations converge to each other.


\begin{figure}[h!]
    \centering
    \includegraphics[scale=.95]{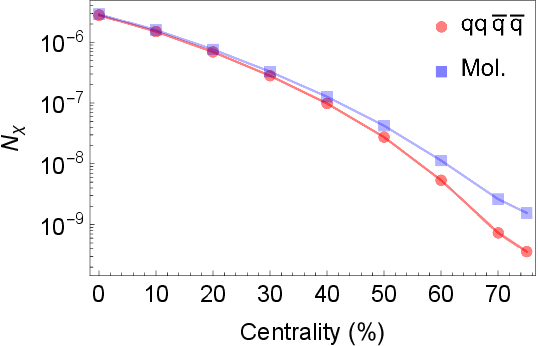}
    \caption{$\chi_{c1}$ multiplicity as a function of the centrality. The curves represent the results obtained with initial conditions associated to the molecular and compact tetraquark interpretations given by Eq.~(\ref{initialconditions}).}
    \label{NYCentrality}
\end{figure}

Finally, in Fig.~\ref{NYCentrality} we present  the $\chi_{c1}$ multiplicity as a function of the centrality, obtained by using Eqs.(\ref{temperaturerelation}),~(\ref{centrality}) and~(\ref{energy}) in (\ref{rateequation}). As expected the $\chi_{c1}(4274)$ final yield decreases as we move from central to peripheral collisions. As before the curves for the molecular and compact tetraquark configurations 
converge to  similar values. 



\section{Concluding Remarks}

In this work we have estimated the  yield of the $\chi_{c1}(4274)$ state in heavy-ion collisions. In our previous work  ~\cite{Britto:2023iso} we found big differences between the thermal cross sections for the production and suppression  of $\chi_{c1}(4274)$. Here we have employed them in the rate equation to determine the $N_{\chi_{c1}}$ evolution in time. We have also added the $\chi_{c1}(4274)$ decay and regeneration terms by means of an effective coupling, estimated from the available experimental data, to account for its small lifetime. We have used the coalescence model to compute the initial multiplicity of the $\chi_{c1}(4274)$, which was treated as a $P-$wave bound state of $D_s\bar D_{s0}$ and also as a compact tetraquark.  As it happened in the case
of other multiquark states, meson molecules are much  more abundant than tetraquarks. Our results indicate  that the combined effects of hadronic interactions, hydrodynamical expansion and the $\chi_{c1}(4274)$ decay strongly affect the final yield. While for tetraquarks  the $\chi_{c1}$  yield grows with respect to the initial value, for molecules it decreases. The final number of $\chi_{c1}$'s is approximately equal for both 
configurations. One can thus conclude that from the multiplicity alone, we cannot determine the intrinsic nature of the $\chi_{c1}(4274)$.    

We have also studied the time evolution of the state $Y^{\prime}(4274)$  proposed in  \cite{zhu2022possible} which is characterized by a smaller width and  a smaller coupling constant obtained via Weinberg compositeness condition. Our results indicate that the  state $Y^{\prime}(4274)$ has a final yield about one order of magnitude higher than that of the $\chi_{c1}$. Hence, if a vertex detector would be  able to accumulate a number of charmed strange mesons as large as $10^4$,  a small number of $Y^{\prime}(4274)$ might be observed. 

As pointed in section~\ref{Formalism}, it is important to remark that the nature of the $\chi_{c1}$ is encoded in the initial condition, fixed from the coalescence model, which implies a different coalescence mechanism for a hadron molecule and for a compact tetraquark. In the analysis of the hadronic interactions happening during the hadron gas phase, we have employed the same coupling for both scenarios, calculated from data. A natural improvement of this work would be  to look at this coupling in more detail. This might be done, for example,  
with  the QCD sum rules techniques.



 \section{ACKNOWLEDGMENTS}

The work of L.M.A. is partly supported by the Brazilian agencies CNPq (Grant Numbers 309950/2020-1, 400215/2022- 5, 200567/2022-5), FAPESB (Grant Number INT0007/2016) and CNPq/FAPERJ under the Project INCT-F\'isica Nuclear e Aplica\c{c}\~oes (Contract No. 464898/2014-5).


\end{document}